\documentclass[preprint,aps,floatfix,showpacs]{revtex4}
\usepackage{graphicx}

\newcommand{\bea}{\begin{eqnarray}}
\newcommand{\eea}{\end{eqnarray}}
\begin{document}
\preprint{\vbox{\hbox{JLAB-THY-04-37} }}
\vspace{0.5cm}
\title{\phantom{x}
\vspace{0.5cm} Decays of Non-strange Negative Parity  Baryons in
the $1/N_c$ Expansion \footnote{This work is dedicated to the memory of Professor Luis Masperi.}}

\author{
J. L. Goity $^{a,b}$ \thanks{e-mail: goity@jlab.org},
\ C. Schat $^{c,d}$  \thanks{e-mail: schat@tandar.cnea.gov.ar}, \ N.
N. Scoccola $^{c,d,e}$
\thanks{e-mail: scoccola@tandar.cnea.gov.ar}}

\affiliation{
$^a$ Department of Physics, Hampton University, Hampton, VA 23668, USA. \\
$^b$  Thomas Jefferson National Accelerator Facility, Newport News, VA 23606, USA. \\
$^c$ Physics Depart., Comisi\'on Nacional de Energ\'{\i}a
At\'omica,
     (1429) Buenos Aires, Argentina.\\
$^d$ CONICET, Rivadavia 1917, (1033) Buenos Aires, Argentina.\\
$^e$ Universidad Favaloro, Sol{\'\i}s 453, (1078) Buenos Aires,
Argentina.}
\date{\today}
\begin{abstract}
The decays of non-strange negative parity baryons via the emission
of single $\pi$ and $\eta$ mesons are analyzed in the framework of
the $1/N_c$ expansion. A basis of spin-flavor operators for the 
partial wave amplitudes is established to  order $1/N_c$ and the unknown effective
coefficients are  determined by fitting to the S- and D-wave
partial widths as provided by the Particle Data Group. A  set of relations
between widths that result at the leading order, i.e. order
$N_c^0$, is given and tested with the available data. 
Up to a few exceptions, a good description of the partial decays widths is already obtained at that order. Because of the rather large
      errors in the empirical input data the next to leading order fit fails to pin down with satisfactory accuracy the  subleading
      effective coefficients.  
The  hierarchy
expected from the $1/N_c$ expansion is  reflected in the results.
\end{abstract}

\maketitle

\section{Introduction}
The  $1/N_c$ expansion has proven  to be a very useful tool for
analyzing the baryon sector. This success  is mostly a consequence
of the emergent contracted spin-flavor symmetry in the large $N_c$
limit \cite{GervaisSakita,DashenManohar}. In the sector of ground
state baryons (identified for $N_c=3$ with the spin 1/2 octet and
spin 3/2 decuplet in the case of three flavors), that symmetry
gives rise to several important relations that hold at different
orders in the  $1/N_c$ expansion
\cite{GroundStates1,Georgi,GroundStates2}. The domain of excited
baryons (baryon resonances) has also been explored in the
framework of the $1/N_c$ expansion
\cite{Goity,PirjolYan,CCGL,GSS,CohenLebed,CaroneGeorgi,CarlsonCarone,PirjolSchat,Stancu}
with very promising results. The analyses carried out so far have
been constrained to states that belong to a definite spin-flavor
and orbital multiplet, i.e. the possibility of mixing of different
such multiplets (so called configuration mixing) has been
disregarded. One good reason for this approximation is that the
resonance data is not sufficient for a full fledged $1/N_c$
analysis that includes those effects. It is also very likely that
such effects are small for dynamical reasons. Indeed, it has been
shown that the only configuration mixings that are not suppressed
by $1/N_c$ factors involve couplings to the orbital degrees of
freedom \cite{GoityMixDecay}. The $1/N_c$ analyses of excited
baryon masses have shown that orbital angular momentum couplings
turn out to be very small \cite{CCGL,GSS}, which is in agreement
with older results in the quark model \cite{CapstickRoberts}. This
strongly suggests that a similar suppression, which is not a
consequence of the $1/N_c$ expansion but rather of QCD dynamics,
also takes place in configuration mixings. Thus, disregarding
configuration mixing is  likely to be a good approximation for the
purpose of phenomenology.  Within such a framework, a few analyses
of excited baryon strong decays have been carried out, namely the decays
of the negative parity SU(6) 70-plet \cite{CaroneGeorgi} and of
the Roper 56-plet \cite{CarlsonCarone}. In the
case of interest in the present work, namely the 70-plet, the analysis in
Ref. \cite{CaroneGeorgi}  used an incomplete basis of
operators at  sub-leading order  in $1/N_c$.   One of the
motivations of the present work is to provide a complete analysis
to ${\cal{O}}(1/N_c)$ for the decays of the non-strange members of
the 70-plet (i.e., the mixed symmetry  20-plet of SU(4)) into
ground state baryons plus a pion or an eta meson. In particular, a
complete basis of effective operators that provide the various S-
and D-wave amplitudes is furnished. It should be advanced that the
lack of a complete basis of operators in \cite{CaroneGeorgi}  does
not affect in a significant way the conclusions of that work. As
discussed later in this work, the fact that the input partial
widths have  large errors implies a rather uncertain determination
of sub-leading effects.

The negative parity 70-plet is the experimentally best established
and known excited baryon multiplet. In particular the S- and
D-wave partial decay widths have been determined from different
data analyses \cite{PDG} with varying degrees of certainty. In
all, these available widths provide sufficient input for the
analysis at ${\cal{O}}(1/N_c)$ pursued in this work. To make the
analysis more conclusive, however, higher precision in the inputs
would be required.

This paper is organized as follows: section II contains the framework for calculating the decays, section III provides the basis of effective operators, section IV presents the results, and finally the conclusions are given in section V.

\section{ Framework for decays  }

In the application of the $1/N_c$ expansion to excited baryons the assumption is made that there is an approximate spin-flavor
symmetry. This assumption is rigorous in the large $N_c$ limit for the ground state baryons. For excited states, however, the symmetry
is broken at order $N_c^0$. This has been found to be the case for the masses of the negative parity baryons, where it was shown \cite{Goity,CCGL} that effective mass operators of  spin-orbit type  produce such a breaking. In the large $N_c$ limit a different scheme becomes rigorously valid \cite{PirjolSchat,CohenLebed}. In the real world with $N_c=3$  there is, however,  plenty of evidence that the dominating spin-flavor breaking in masses are the hyperfine effects that are order $1/N_c$, while the spin-orbit effects are substantially smaller
\cite{CCGL,GSS}. Thus, a scheme based on an approximate  spin-flavor symmetry is very convenient for  phenomenological purposes.

The excited baryons are therefore classified in multiplets of the $O(3)\times SU(2N_f)$ group. $O(3)$
corresponds to spatial rotations and $SU(2N_f)$ is the spin-flavor group where $N_f$ is the number of flavors
being considered, equal to two in the present work. The ground state baryons, namely the $N$ and $\Delta$ states,
belong to the $({\bf 1},{\bf 20_S})$ representation, where the $\bf 20_S$ is the  totally symmetric
representation of $SU(4)$. The
negative parity baryons considered here belong instead  to the $({\bf 3},{\bf 20_{MS}})$ representation,  where  $\bf 20_{MS}$
is the mixed symmetric
representation of $SU(4)$. For general $N_c$ the spin-flavor representations involve, in the Young tableaux
language,  $N_c$ boxes  and are identified with  the totally symmetric and the
mixed-symmetric representations of type $(N_c-1,1)$  for ground  and excited negative
parity states respectively. Since in the mixed symmetric spin-flavor representation one
box of the Young tableaux is distinguished,  such a box associated with  the \lq\lq excited
quark\rq\rq \   in the baryon. In a similar fashion, and without any loss of generality it
is possible to distinguish one box in the ground state multiplet as well.  This is a very
convenient  procedure    that has been used repeatedly in previous works.  The  spin and
isospin quantum numbers of the distinguished box will be denoted with lower cases, and the
corresponding quantum numbers of the rest of $N_c-1$ boxes (which are in a totally symmetric
representation of $SU(4)$  and form the so  called  \lq\lq core\rq\rq ~ of the large $N_c$
baryon) after they are coupled to eigenstates of spin and isospin will be denoted by $S_c$
and $I_c$ respectively.  Notice  that $S_c=I_c$ for totally symmetric representations of
$SU(4)$. For a given core state, the coupling of the excited quark gives eigenstates of
spin and isospin:
\begin{eqnarray}
\mid S,  S_3 ; I , I_3 ; S_c\rangle &=&\sum_{s_3,i_3} \langle S_c, S_3-s_3; \frac{1}{2}, s_3  \mid S, S_3\rangle
\langle I_c=S_c, I_3-i_3; \frac{1}{2}, i_3  \mid I, I_3 \rangle \nonumber\\
&\times&
   \mid S_c ,  S_3-s_3 ; I_c=S_c, I_3-i_3 \rangle \mid \frac{1}{2}, s_3 ; \frac{1}{2}, i_3   \rangle
\end{eqnarray}
These states are not in an irreducible representation  of $SU(4)$, as they are not in an irreducible representation of the permutation group. The totally symmetric states are given by:
\begin{eqnarray}
\mid S, S_3 ; I , I_3 \rangle_{\mbox{\bf S}} &=&\sum_{\eta=\pm \frac{1}{2}}
C_{\mbox{\bf S}}(S,\eta) \mid S , S_3 ; I , I_3 ; S_c=S+\eta\rangle,
\end{eqnarray}
where
\begin{eqnarray}
C_{\mbox{\bf S}}(S,\pm \frac{1}{2})&=&  \sqrt{\frac{(2 S + 1 \mp 1 ) (N_c+ 1 \pm (2 S+ 1))}{2 N_c ( 2 S + 1)} } \ ,
\end{eqnarray}
while  the mixed symmetric ({\bf MS}) states   $(N_c-1,1)$ are given by:
\begin{eqnarray}
\mid S, S_3,\; I \; I_3\rangle_{\mbox{\bf MS}} &=&\sum_{\eta=\pm \frac{1}{2}}
C_{\mbox{\bf MS}}(I,S,\eta) \mid S, S_3 ; I , I_3 ; S_c=S+\eta \rangle ,
\end{eqnarray}
where
\begin{eqnarray}
C_{\mbox{\bf MS}}(I,S, \pm \frac{1}{2})&=&\left\{
    \begin{array}{c}
      1~~{\rm if}~~I=S \pm 1   \\
      0~~{\rm if} ~~ I=S \mp 1 \\
      \pm \sqrt{\frac{(2 S + 1 \pm 1 ) (N_c+ 1 \pm (2 S+ 1))}{2 N_c ( 2 S + 1)} } ~~{\rm if} ~~I=S
    \end{array}\right.
\end{eqnarray}
Finally, upon coupling the orbital degrees of freedom, the excited baryons in the
$({\bf 3},{\bf {MS}})$ representation are given by:
\begin{equation}
\mid J , J_3 ; I , I_3 ; S \rangle_{\mbox{\bf MS}} =
\sum_m \langle   1, m ; S , J_3 - m \mid J, J_3\rangle  \mid 1 , m \rangle
\mid S, J_3 - m ;I , I_3 \rangle_{\mbox{\bf MS}}.
\end{equation}
For $N_c=3$ the states are displayed in  Table I along with their
quantum numbers, masses, decay widths and branching ratios.

 Note that there are two sets of $N^*$ states each consisting of two states with the same spin and isospin.
 The physical states are admixtures of such states, and are given by:
\bea
\left(
    \begin{array}{c}
        N^*_{J}  \\
        N^{*'}_{J}
    \end{array}
\right)
&=&
\left(
    \begin{array}{rr}
        \cos{\theta_{2 J }}  & \sin{\theta_{2 J }}  \\
            - \sin{\theta_{2 J }}  &  \cos{\theta_{2 J }}
    \end{array}
\right)
\left(
    \begin{array}{c}
        ^2N^*_{J}  \\
        ^4N^*_{J}
    \end{array}
\right) \ ,
\eea
where $J=\frac{1}{2}$ and $\frac{3}{2}$, $N^{*(')}_J$ are mass eigenstates, and
the two mixing  angles can be constrained to be in the interval $[0 , \pi)$.
Here  the notation $^{2S+1}N^*_{J}$ has been used.

The possible strong decays with emission of a single $\pi$ or
$\eta$ meson (including those in the $\eta-$channel that turn out
to be kinematically forbidden) are shown in Table I. Only S- and
D- wave decays are considered \cite{foot}. 
The effective amplitudes for
the emission of a pseudoscalar meson have the most general form:
\bea
M^{[\ell_P,I_P]}(\vec{k}_P)&=&(-1)^{\ell_P} \ \sqrt{2 M_{B*}}\
Y^*_{\ell_P, m_P}(\hat{k_P})\sum_{\mu,\alpha}
\langle \ell_P , m_P ; I_P , {I_P}_3 \mid P^{[\ell_P, I_P]}_{[\mu, \alpha]} \mid 0
\rangle \times
\nonumber\\
& & \qquad \qquad
\ _{\mbox{\bf S}} \langle  S , S_3 ; I , I_3 \mid B^{[\ell_P, I_P]}_{[-\mu,-\alpha]}
\mid   J^*, J^*_3 ; I^*, I^*_3  ; S^*  \rangle_{\mbox{\bf MS}},
\eea
where $\ell_P$ and $I_P$ are the orbital angular
momentum and isospin of the pseudoscalar meson, and $\mu$ and $\alpha$ the corresponding
projections.
$P^{[\ell_P, I_P]}_{[\mu, \alpha]}$ is a mesonic operator
that creates the final state pseudoscalar  and  $B^{[\ell_P, I_P]}_{[-\mu,-\alpha]}$
is a baryonic operator that transforms the
initial excited baryon into a ground state baryon.
Here, the partial wave is projected onto a meson momentum eigenstate
with momentum $\vec{k}_P$. The factor $\sqrt{2 M_{B*}}$, where $M_{B*}$
is the mass of the excited baryon, has been added for convenience. Since
as shown later,
the dynamics of the decay can be encoded in effective coefficients,
the mesonic operator matrix elements can be chosen to be:
\begin{equation}
\langle \ell_P , m_P ; I_P , {I_P}_3 \mid P^{[\ell_P, I_P]}_{[\mu, \alpha]} \mid 0 \rangle =
\sqrt{(2 \ell_P + 1)( 2 I_P + 1)} \ \delta_{\mu\, m_P} \delta_{\alpha \,{I_P}_3} .
\end{equation}

The baryonic operator admits an expansion in $1/N_c$ and has the general form:
\begin{equation}
B^{[\ell_P, I_P]}_{[\mu, \alpha]} =
\left(\frac{k_P}{\Lambda}\right)^{\ell_P}\sum_q \, C_q^{[\ell_P, I_P]}(k_P)
\left( B^{[\ell_P, I_P]}_{[\mu, \alpha]} \right)_q,
\label{exp}
\end{equation}
where
\begin{equation}
\left( B^{[\ell_P, I_P]}_{[\mu, \alpha]} \right)_q = \sum_m
\langle 1, m ; j, j_z  \mid \ell_P, \mu \rangle \
\xi^1_m \ \left( {\cal{G}}^{[j, I_P]}_{[j_z,\alpha]} \right)_q,
\end{equation}
and the factor  $\left(\frac{k_P}{\Lambda}\right)^{\ell_P}$ is included to take
into account the chief meson momentum dependence of the partial wave. The scale $\Lambda$ is chosen in what follows to be 200 MeV.
Here, $\xi^1_m$ is an operator that produces a transition from the triplet to the singlet $O(3)$ state,
and $\left( {\cal{G}}^{[j, I_P]}_{[j_z,\alpha]} \right)_q$ is a spin-flavor operator that produces the transition from the
mixed-symmetric to the symmetric $SU(4)$ representation. The label  $j$ denotes  the spin of the spin-flavor operator,
and as it is clear, its isospin coincides with the isospin of the emitted meson.
As it was already mentioned, the dynamics of the decay is encoded in the effective dimensionless coefficients $C_q^{[\ell_P, I_P]}(k_P)$.

The terms in the right hand side of Eq.(\ref{exp}) are ordered in powers of $1/N_c$.
As it has been explained in earlier publications \cite{GroundStates1}, the order in $1/N_c$
is determined by the spin-flavor operator. For an $n$-body operator, this order is given by
\begin{equation}
\nu=n-1-\kappa,
\end{equation}
where $\kappa$ is equal to zero for incoherent operators and can
be equal to one or even larger for coherent operators. More details can be found in the following section
where a basis of operators ${\cal{G}}$ is explicitly built.

With the definition of effective operators used in this work, all  coefficients
$C_q^{[\ell_P, I_P]}(k_P)$ in Eq.(\ref{exp}) are of zeroth order in $N_c$.
The leading order of the
decay amplitude is in fact $N_c^0$ \cite{GoityMixDecay}.
At this point it is important to comment on the momentum dependence of
the coefficients. The  spin-flavor breakings  in the masses, of both excited
and ground state baryons,  give rise to different values of the momenta $k_P$.
In the expansion where the spin-flavor breaking is treated as small,
$k_P=\overline{k_P}+\delta k_P$, where $\overline{k_P}$ is the spin-flavor symmetry
limit value, and $ \delta k_P$  is taken as small and expanded upon.
Only the momentum dependence of the coefficients associated with leading order
operators should be included if one takes  $ \delta k_P$  to be of the same order
as $1/N_c$ corrections.  The goal would be to treat these corrections in a model
independent fashion. In an operator analysis, they require the consideration of
reducible effective operators that are the product of a mass operator that gives
the spin-flavor breaking mass shifts (e.g. of  the ground state baryons) times the
leading order operator to which the coefficient is associated with.
These corrections should be considered to be of the  same order as the
subleading in $1/N_c$ corrections given by irreducible operators that are
analyzed below in full detail. In principle,  this could be achieved with
no difficulties in a world where $N_c$ is large.
The problem is that for $N_c=3$ the number of independent amplitudes
for the decays here under consideration is  essentially the same as
the number of irreducible operators that appear in the analysis to order $1/N_c$.
There is, therefore,  no room to separate the momentum dependence effects in the
coefficients from the $1/N_c$ effects due to irreducible operators.
Thus,  one can adopt the scheme followed in this paper where the only
momentum dependence assigned to the coefficients is the
explicitly shown factor    $\left(k_P/\Lambda\right)^{\ell_P}$ that
takes into account the chief momentum dependence of the corresponding
partial wave,  and the rest of the dependence is then encoded in the coefficients
of the sub-leading operators. The other possibility is to model the momentum dependence,
as it was done in \cite{CarlsonCarone}, with a profile function motivated
for instance by a quark model, in such a way that less of the momentum dependence is
absorbed by subleading operators.   Both ways of proceeding are  equally valid.

Using the standard definition for the decay width
and averaging over the initial- and summing over
the corresponding final-baryon spins and isospins
,  the decay width for each $[\ell_P, I_P]$
channel is given by
\begin{equation}
\Gamma^{[\ell_P,I_P]}
= f_{ps} \
\frac{|\sum_q C_q^{[\ell_P, I_P]}\
{\cal B}_q(\ell_P,I_P,S,I,J^*,I^*,S^*)|^2}
{\sqrt{(2 J^* + 1)(2 I^*+1)}},
\label{width}
\end{equation}
where the phase space factor $f_{ps}$ is
\begin{equation}
f_{ps} =\frac{k_P^{1+2\ell_P}}{8 \pi^2 \Lambda^{2\ell_P}} \frac{M_{B^*}}{M_B}
\label{fsp}
\end{equation}
and
${\cal B}_q(\ell_P,I_P,S,I,J^*,I^*,S^*)$ are reduced matrix elements defined
via the Wigner-Eckart theorem as follows:
\bea
& & _{\mbox{\bf S}} \langle  S , S_3 ; I , I_3 \mid
\left( B^{[\ell_P, I_P]}_{[m_P,I_{P_3}]}\right)_q
\mid   J^*, J^*_3 ; I^*, I^*_3  ; S^*  \rangle_{\mbox{\bf MS}}
= \frac{ (-1)^{\ell_P - J^* + S + I_P - I^*+I} }{\sqrt{(2 S + 1)(2 I + 1)}}
  \qquad \qquad \nonumber \\
&\times & \qquad
\langle \ell_P , m_P; J^*, J^*_3 \mid S, S_3  \rangle  \langle I_P, {I_P}_3 ; I^*, I^*_3 \mid I=S , I_3  \rangle
\ {\cal B}_q(\ell_P,I_P,S,I,J^*,I^*,S^*) \ .
\eea
These reduced matrix elements
can be easily calculated in terms of the reduced matrix elements of the spin-flavor operators,
namely
\bea
{\cal B}_q(\ell_P,I_P,S,I,J^*,I^*,S^*) &=&
 (-1)^{j+ J^* + \ell_P + J + 1} \sqrt{(2 J^* + 1)(2 \ell_P +1)} \
\left\{ \begin{array}{ccc}
     J^*   & S^*     & 1   \\
     j     & \ell_P     & S
\end{array} \right\} \nonumber \\
& \times &
\qquad _{\mbox{\bf S}} < S ;I  || \left( {\cal G}^{[j,I_P]} \right)_q || S^*; I^* >_{\mbox{\bf MS}},
\label{red},
\eea
where, due to the fact that  the unknown dynamics can be included in the effective coefficients,
the operators $\xi^1_m$ can be chosen such that their matrix elements are simply given by
\begin{equation}
\langle 0 \mid \xi^1_{m'} \mid 1\;m \rangle=- \sqrt3 \ \delta_{m\, m'}.
\end{equation}
Note that in the present case, where $\ell_P$ can be 0 or 2 only, Eq.(\ref{red}) implies that the
spin-flavor operators can carry spin $j$ that can be 1, 2 or 3.

\section{ Basis of operators}

The construction of a basis of spin-flavor operators follows
similar lines as in previous works on baryon masses. The
spin-flavor operators considered in the present paper must
connect a mixed-symmetric with a symmetric representation.
Generators of the spin-flavor group acting on the states
obviously do not produce such connection. However, generators
restricted to act on the excited quark or on the core of
$N_c-1$ quarks can do this. The spin flavor operators can,
therefore,  be represented by products of generators of the
spin-flavor group restricted to act either on the excited
or on the core states.
 In the following the generators acting on the core are denoted by $S_c$,
$G_c$, $T_c$, and the ones acting on the excited quark by $s$, $g$, $t$. The
generators $G_c$  are known to be coherent operators, while all the rest are
incoherent. In order to build a basis of operators for the present problem one
has to consider products of such generators with the appropriate couplings of
spins and isospins. The $n$-bodyness ($n$B) of an operator is given by the number of
such factors, and the level of coherence of the operator is determined by how
many factors $G_c$ appear in the product. It should be noticed that  in the physical case where $N_c=3$, only operators of at most 3B have
to be considered. Still, to order $1/N_c$ there is a rather long list of
operators of given spin $j$ and isospin $I_P$. This list can be drastically
shortened by applying several reduction rules. The first rule is that the
product of two or more generators acting on the excited quark can always be
reduced to the identity operator or to a linear combination of  such  generators.
The second set of rules can be easily
derived for products of operators whose  matrix elements are taken  between a
mixed-symmetric and a symmetric representation. These reduction rules are as
follows for 1B  up to 3B operators (here $\lambda$ represents generators acting on the excited quark and
$\Lambda_c$ represent  generators acting on the core):
\bea
\lambda&=&-\Lambda_c \nonumber\\
(\Lambda_c)_1 (\Lambda_c)_2
&=&- \lambda_1 (\Lambda_c)_2- \lambda_2 (\Lambda_c)_1 + {\rm 1B~ operators}.
\eea
Therefore, only the following types of operators should be considered
\begin{eqnarray}
&\mbox{1B}&  \qquad \lambda \nonumber \\
&\mbox{2B}&  \qquad \frac{1}{N_c} \ \lambda_1 \ (\Lambda_c)_2
\label{123B}\\
&\mbox{3B}&  \qquad \frac{1}{N_c^2} \ \lambda_1 \ (\Lambda_c)_2 \ (\Lambda_c)_3 \nonumber
\end{eqnarray}

It is convenient to make explicit the transformation
properties of each basic operator under spin $j$ and isospin $t$.
In what follows  the notation $O^{[j,t]}$ is used  to indicate that the
operator $O$ has spin $j$  and isospin $t$. It is easy to see that
\begin{eqnarray}
\lambda^{[j,t]} &=& s^{[1,0]},\ t^{[0,1]}, \ g^{[1,1]} \nonumber \\
(\Lambda_c)^{[j,t]} &=& (S_c)^{[1,0]},\ (T_c)^{[0,1]},
\ (G_c)^{[1,1]} \label{transbasic}
\end{eqnarray}
For decays in the $\eta-$channels the spin-flavor operators  transform as $[j,0]$ while
for decays in the pion channels they should transform as $[j,1]$, where in both
cases $j=1,2,3$. Knowing the transformation properties of each
basic operator given in Eq.(\ref{transbasic}) it is easy to
construct products of the forms given in  Eqs.(\ref{123B}) with the desired spin and isospin. An example of a 2B operator that
transforms as $[2,1]$ is
\begin{equation}
 (g G_c)^{[2,1]}_{[\mu,\alpha]} = < 1, m; 1, m' | 2, \mu >
< 1, a; 1, a' | 1, \alpha > \  g_{m a} \ (G_c)_{m'a'},
\end{equation}
where  the spin and isospin projections (e.g. $m$,
$m'$, $a$, etc) are considered to be spherical.

Thus, the 1B operators that contribute to a given $[j,t]$ are just
those $\lambda^{[j,t]}$ given in Eq.(\ref{transbasic})
which have the proper spin and isospin quantum numbers. Similarly,
the possible 2B operators are given by the products
$\lambda_1^{[j_1,t_1]}
(\Lambda_c)_2^{[j_2,t_2]}$ coupled to the required
$[j,t]$ by means of the conventional Clebsch-Gordan (CG)
coefficients.

In order to construct all possible 3B operators, it is convenient to note
that there are additional reduction rules for the products
$(\Lambda_c)_2 (\Lambda_c)_3$. These rules apply to  matrix elements of products of generators between states in the symmetric representation \cite{ DashenManohar}. The starting
point is to consider all possible products of two core operators.
Since one is interested in keeping contributions of at most order
$1/N_c$, and because these products will appear only in  3B
operators, at least one of the operators must be a $G_c$. Thus, the list of 2B core operators to be considered is:
\begin{eqnarray}
\left( T_c \ G_c \right)^{[1,t]} \ , \
\left( G_c \ T_c \right)^{[1,t]} \ , \
\left( S_c \ G_c \right)^{[j,1]} \ , \
\left( G_c \ S_c \right)^{[j,1]} \ , \
\left( G_c \ G_c \right)^{[j,t]},
\end{eqnarray}
where $j,t=0,1,2$. Using transformation properties of the CG
coefficients and dropping all the combinations leading to
commutators (because the basic operators are  generators of the
algebra, the commutators are proportional to another basic
operator), only the following products remain
\begin{eqnarray}
( \{ T_c, G_c \} )^{[1,0]} \qquad &,& \qquad ( [ T_c, G_c ] )^{[1,1]} \qquad , \qquad ( \{ T_c, G_c \} )^{[1,2]} \\
( \{ S_c, G_c \} )^{[0,1]} \qquad &,& \qquad ( [ S_c, G_c ]
)^{[1,1]} \qquad , \qquad ( \{ S_c, G_c \} )^{[2,1]}.
\end{eqnarray}
\begin{eqnarray}
(G_c \ G_c)^{[0,0]} \ , \
(G_c \ G_c)^{[0,2]} \ , \
(G_c \ G_c)^{[1,1]} \ , \
(G_c \ G_c)^{[2,0]} \ , \
(G_c \ G_c)^{[2,2]},
\end{eqnarray}
where  $( [ T_c, G_c ] )^{[1,1]} $ denotes $ (T_c G_c)^{[1,1]} - (G_cT_c)^{[1,1]}
                      = \langle 1 a 1 b | 1 d \rangle ( T_c^{a}  G_c^{mb} - G_c^{ma} T_c^{b} )
                      =  \langle 1 a 1 b | 1 d \rangle  \{ G_c^{ma} , T_c^{b} \}$, etc..
Using now the reduction relations \cite{DashenManohar},  it
can be shown that several of these eleven  operators can be
eliminated.  The final list of independent products of two core
operators turns out to be
\begin{eqnarray}
( \{ T_c, G_c \} )^{[1,2]} \ , \
( [ S_c, G_c ]   )^{[1,1]} \ , \
( \{ S_c, G_c \} )^{[2,1]} \ , \
(G_c \ G_c)^{[2,2]}
\end{eqnarray}
By coupling any of these operators with one of the excited core
operators $\lambda_1^{[j_1,t_1]}$ (see
Eq.(\ref{transbasic})) to the required $[j,t]$
all the possible 3B operators are obtained.

Using this scheme to couple products
of generators it is straightforward to
construct lists of operators with spin
1, 2 and 3 and isospin 0 and 1.
Further reductions result from the fact that
not all the resulting  operators are linearly independent
up to order $1/N_c$. The determination of  the final set of independent
operators for each particular decay channel is more laborious. This is achieved by coupling
the resulting spin-flavor operators with  the
$\xi^{[1,0]}$ orbital transition operator to the corresponding
total spin and isospin and by explicitly calculating  all the
relevant matrix elements.
In this way  operators that are linearly dependent at the corresponding order in the $1/N_c$ expansion can be eliminated.
The resulting  basis of independent operators
$\left( O^{[\ell_P, I_P]}_{[m_P,I_{P_3}]}\right)_q$ is shown in Table II, where for
simplicity the corresponding spin and isospin projections have been omitted. Their
 reduced matrix elements are given in Tables III through VI. In the bottom rows of these
tables   normalization coefficients
$\alpha^{[\ell,I_P]}_q$ are displayed. These coefficients are
used to define normalized basis operators
such that, for $N_c=3$, their largest
reduced matrix element is equal to one for order $N_c^0$ operators
and equal to 1/3 for order $1/N_c$ operators. Thus,
\begin{equation}
\left( B^{[\ell_P, I_P]}_{[m_P,I_{P_3}]}\right)_q =
\alpha^{[\ell,I_P]}_q  \left( O^{[\ell_P, I_P]}_{[m_P,I_{P_3}]}\right)_q
\end{equation}
furnishes the list of basis operators normalized according to the $1/N_c$ power counting.

 \section{Results}

 The  different  S- and D-wave partial widths used in the analysis are the ones provided by the Particle Data
 Group \cite{PDG}.  The values for the widths and branching ratios are taken as the ones  indicated there as
 \lq\lq our estimate\rq\rq, while the errors are determined from the corresponding ranges.
 The total widths and branching fractions are given in Table I, while in Table VIII the partial widths calculated
 from those values are explicitly displayed. The entries  in Table I indicated  as unknown reflect
 channels for which no width is provided by the Particle Data Group or where the authors consider that the input is unreliable,
 such as in the  $\pi \Delta$ decay modes of the $N(1700)$ and $N(1675)$ and the D-wave $\eta-$ decay modes.
 At this point it is important to
 stress the marginal precision of the data for the purposes of this work. This work performs an analysis at
 order $1/N_c$, which means that the theoretical error  is order $1/N_c^2$.  This implies that amplitudes are
 affected by a theoretical uncertainty at the level of 10\%. Thus, in order  to pin down the coefficients of the
 subleading operators, the widths provided by the data should not  be affected by errors larger than about 20\%.
 As shown in Tables I and VIII, the experimental errors are in most entries 30\% or larger. In consequence, the
 determination of the subleading effective coefficients is affected by large errors as the results below show.

Before presenting the results of the fits,  it is convenient to derive some parameter
independent relations that can be obtained to leading order.  These relations serve as a test
of the leading order approximation.  Since at this order there are only four coefficients and two angles to be
fitted, and there are a total of twenty partial widths (excluding all D-wave $\eta$ channels
but including kinematically forbidden $\eta$-channel decays), there are fourteen  independent
parameter free relations that can be derived. These relations are more conveniently written in
terms of reduced widths, i.e., widths where the phase space factor $f_{sp}$ (see Eq.(\ref{fsp}))
has been removed and denoted
here by $\tilde \Gamma$. Considering the S-wave decays in the $\pi$ mode, there are six decays
and three parameters in the fit. Thus, three parameter free relations must follow. These relations and the corresponding comparison with experimental values read:
{\small{
\begin{equation}
\left.
\begin{array}{cccccccc}
 &{\begin{array}{c}\tilde{\Gamma}_{N(1535)\to\pi N}\\
 +\tilde{\Gamma}_{N(1650)\to\pi N}
 \end{array} }&:&{\begin{array}{c} \tilde{\Gamma}_{N(1520)\to\pi \Delta}\\+
 \tilde{\Gamma}_{N(1700)\to \pi \Delta} \end{array} }
 &:&\tilde{\Gamma}_{\Delta(1620)\to\pi N}&:&\tilde{\Gamma}_{\Delta(1700)\to\pi\Delta}\\
{\rm Th.}& 1        &:& 1                            &:&0.17                 &:& 0.42 \\
{\rm Exp.}&       1 &:& {\rm unknown}     &:&0.19\pm 0.07 &:& 0.62\pm 0.33
\end{array}
\right. .
\end{equation}}}
Within the experimental errors the relations are satisfied. They can be used to give  a leading order prediction
for the unknown S-wave width: $\Gamma_{N(1700)\to \pi \Delta} = 160 \pm 40 \,{\rm MeV}$. The following expression
for the $\theta_1$ mixing angle also follows:
\begin{equation}
\frac{\tilde{\Gamma}_{N^(1535)\to\pi N}-\tilde{\Gamma}_{N^(1650)\to\pi
N}}{\tilde{\Gamma}_{N^(1535)\to\pi N}+\tilde{\Gamma}_{N^(1650)\to\pi N}}=
 \frac{1}{3}\left[ \cos
(2\theta_1)-\sqrt{8} \sin(2\theta_1)\right].
\label{thetauno}
\end{equation}
Using the empirical values the angle that results from
this relation is: $\theta_1= 1.62\pm 0.12$ or $0.29\pm 0.11$.
 One of these angles, namely the latter,
turns out to be close to the angle obtained in the next to leading order fit.

In a similar fashion, relations
involving D- wave decays in the $\pi$ mode  can be obtained. There are in this case four parameters to fit and
eleven partial widths. Thus, seven parameter free relations can be obtained. In general these relations
are quadratic and/or involve some of the unknown decay widths. However, the following testable three linear relations
can be obtained
\begin{eqnarray}
2\ \tilde{\Gamma}_{\Delta(1620)\to\pi \Delta} + \tilde{\Gamma}_{\Delta(1700)\to\pi \Delta}&=&
8\ \tilde{\Gamma}_{\Delta(1700)\to\pi N} + \frac{15}{4}\ \tilde{\Gamma}_{N(1675)\to\pi N} \nonumber \\
{\rm Exp.} \qquad \qquad \qquad \qquad \qquad 5.24 \pm 1.95 \qquad \qquad &=& \qquad \qquad 2.19 \pm 0.62
\end{eqnarray}
\begin{eqnarray}
\frac{2}{9}\left(\tilde{\Gamma}_{\Delta(1535)\to\pi \Delta} + \tilde{\Gamma}_{N(1650)\to\pi \Delta} \right)+
\frac{20}{3}\ \tilde{\Gamma}_{\Delta(1620)\to\pi \Delta}&= &
 16\ \tilde{\Gamma}_{\Delta(1700)\to\pi N} + 15\ \tilde{\Gamma}_{N(1675)\to\pi N} \nonumber \\
{\rm Exp.}  \!\!\!\! \qquad\qquad\qquad \qquad \qquad 16.87 \pm 6.46 \qquad \qquad &=& \qquad \qquad 6.49 \pm 1.65
\end{eqnarray}
\begin{eqnarray}
\frac{1}{36}\left(\tilde{\Gamma}_{\Delta(1520)\to\pi N} + \tilde{\Gamma}_{N(1700)\to\pi N} \right)+
\frac{5}{12}\ \tilde{\Gamma}_{\Delta(1620)\to\pi \Delta}& =&
 \tilde{\Gamma}_{\Delta(1700)\to\pi N} +  \tilde{\Gamma}_{N(1675)\to\pi N} \nonumber \\
{\rm Exp.} \qquad \qquad\qquad \qquad \qquad \qquad 1.07\pm 0.40 \qquad \qquad &=& \qquad \qquad 0.42 \pm 0.11
\end{eqnarray}

These relations are not well satisfied
by the empirical data. Thus, one can anticipate the need for next to leading corrections in order to have  a correct description of some D-wave  decays. In the case
of S-wave $\eta-$ mode decays, there
are two parameters to fit (since there
is no dependence on the mixing angle
$\theta_3$), and three possible
decays. However, one of the decays is
kinematically forbidden, namely the
$\Delta(1700)\to \eta \Delta$.  It is
also possible to express the mixing
angle $\theta_1$ in terms of the
reduced S-wave widths:
\begin{equation}
\frac{\tilde{\Gamma}_{N^(1535)\to\eta N}-\tilde{\Gamma}_{N^(1650)\to\eta
N}}{\tilde{\Gamma}_{N^(1535)\to\eta N}+\tilde{\Gamma}_{N^(1650)\to\eta N}}=
-\frac{1}{3}\left[ \cos
(2\theta_1)-\sqrt{8} \sin(2\theta_1)\right].
\end{equation}
Using the empirical values the angle that results from
this relation is: $\theta_1= 1.26\pm 0.14$ or $0.65\pm 0.14$.
Although these results seem not to be far from the results derived from Eq.(\ref{thetauno}),
the leading order fit discussed below indicates a poor description of the ratio of the $N(1650)\to \eta N$
to $N(1535)\to \eta N$ widths. This is due to the high sensitivity of this ratio to the mixing angle
$\theta_1$.
Indeed, it is necessary to include some $1/N_c$ effects as discussed below to arrive at a good description of the $\eta$-modes
together with the other modes, and in this case the resulting angle is $0.39\pm 0.11$.

   In Table VII the results of several fits are displayed. In these fits the decay amplitudes are expanded 
   keeping  only the terms that correspond to the order in $1/N_c$ of the fit. Similarly, when performing the fits
   the  errors have been taken to be equal or larger than the expected accuracy of the fit ($30 \%$
   to the LO fits and about $10\%$ for the NLO ones).

   The first LO fit only
   considers the S-wave $\pi$-modes. As expected, the values for the  mixing angle $\theta_1$ resulting from this fit turn  out to
   be equal to the ones obtained through the relation given by  Eq.(\ref{thetauno}). Notice that $\theta_3$ also has a
   two fold ambiguity at this order. The second leading order fit includes the D-waves and $\eta$-modes. The
   angles remain within errors equal to the ones from the first fit.  Table VIII shows that the  $N(1535)\to
   \eta N$  width results to be a factor four smaller than the empirical one (this having, however, a rather
   generous error). In the D-waves, several widths involving decays with a $\Delta$ in the final state are also
   too small. The S-wave $\pi$-modes are well fitted and there is no real need for NLO  improvement. In the
   D-wave decays in the $\pi$ channel there are two leading order  operators that contribute.  The fit shows
   that the 1B operator has a coefficient whose magnitude  is a factor two  to three larger than that of the
   coefficient of the 2B operator.  The 1B D-wave  operator  as well as  the 1B S-wave operator $O^{[0,1]}_1$
   stem from   the 1B coupling of the pion via the axial current. Such a coupling naturally occurs as the
   dominant coupling in the chiral quark model \cite{Manohar:1983md}. Thus, the result of the D-wave fit might suggest that such a
   mechanism is dominant over  other mechanisms that give rise to  the 2B operator   $O^{[2,1]}_6$.

   The NLO fit involves  a rather large number of effective constants. In addition to the four effective constants
  and the two mixing angles that appear at LO, there are ten new effective constants, three of them in the S-wave pion channels,
  six in the D-wave pion channels, and one in the S-wave eta channels. Since there are only sixteen  data available,
  some operators must be discarded for the purpose of the fit. It is reasonable to choose to neglect
  the 3B operators and the subleading S-wave operator for $\eta$ emission. The NLO fit has been carried
  out by demanding that the LO coefficients are not  vastly different from their values obtained in the LO fit.
  This demand is reasonable if the assumption is made that the $1/N_c$ expansion makes sense. The fact that the
  NLO coefficients do not have unnaturally large values with respect to the scale set by the LO fit indicates the consistency of the assumption. This
  clearly is no proof, however, that the $1/N_c$ expansion is working. As mentioned earlier, the chief
  limitation here is due to the magnitude of the errors in the inputs.   This leads to results for the NLO
  coefficients being affected with rather large uncertainties. Indeed, no clear NLO effects can be pinned down,
  as most NLO coefficients are no more than one standard deviation from zero.  Because  the number of
  coefficients is approximately equal to the number of inputs, there are important correlations between  them.
  For instance, the S-wave NLO coefficients are very correlated with each other and  with  the angle $\theta_3$.
  For the S-waves the LO fit is already excellent, and therefore nothing significantly  new is obtained by including
  the NLO corrections.
  Correlations are smaller for the D-wave coefficients. As mentioned before, here the LO fit has room for
  improvement, and thus the NLO results are more significant than in the case of the S-waves. Still, no clear
  pattern concerning the NLO corrections is observed. One interesting point, however,  is that without any significant
  change in the value of $\theta_1$ the $\eta$-modes are now well described. The reason for this is that in the
  LO fit the matrix elements of the operator $O_1^{[0,0]}$ were taken to zeroth order in $1/N_c$, while in the
  NLO fit the $1/N_c$ terms are included. These subleading corrections enhance the amplitude for the
  $^2N^*_{1/2}$ and suppress the amplitude for the $^4N^*_{1/2}$. This along with an increment in the coefficient
  brings the fit in line with the empirical widths. One important point is that at NLO the two fold ambiguity in
  $\theta_1$ that results at LO is eliminated. The smaller mixing angle turns out to be selected. The angle
  $\theta_3$ remains ambiguous and close to the values obtained in the LO fits. It should be noticed
  that the present values of both mixing angles are somewhat different
  from the values $\theta_1=0.61, \ \theta_3=3.04$  obtained in other analyses \cite{CaroneGeorgi,CapstickRoberts,lll}.
  Finally, a clear manifestation
  of the lack of precision in the data is the ratio between the errors affecting the coefficients of NLO  versus
  those of LO operators. In a situation where the data would have precision of the order of NNLO corrections,
  that ratio should be $\simeq N_c$. In most cases  the ratios turn out to be   much larger than that,  as Table
  VII shows.

\section{Conclusions}

In this work the method of  analysis of excited baryon decays in the $1/N_c$ expansion has been presented. It is
limited to the situation where configuration mixings are neglected, and therefore the baryon states are taken to
belong to a single multiplet of $O(3)\times SU(4)$. The application to the decays of the negative parity baryons
illustrates the method. A basis of effective operators,  in which the S- and D-wave amplitudes are expanded, was
constructed to order $1/N_c$. All dynamical effects are then encoded in the effective coefficients that enter in
that expansion.

The application to the decays addressed here shows that a
consistent description within the $1/N_c$ expansion is possible.
Indeed, up to the relatively poor determination of $1/N_c$
corrections that results from the magnitude of the errors in the
input widths, these corrections are of natural size. A few clear
cut observations can be made. The most important one is that the
S-wave $\pi$- and $\eta$-channels are well described by the
leading order operators (one for each channel) provided one
includes  the contributions subleading in
$1/N_c$ in the matrix elements for the $\eta$-decays. The mixing angle $\theta_1$ is then determined by these
channels up to a twofold ambiguity, which is lifted when all
channels are analyzed at NLO. The angle $\theta_3$ is also
determined up to a two fold ambiguity at LO. The ambiguity remains
when the NLO is considered.   The LO results also indicate the
dominance of  the 1B effective operators that have the structure
that would result from a chiral quark model \cite{Manohar:1983md}. This is explicitly
seen in the D-wave channels where the LO 2B operator turns out to
be suppressed with respect to the 1B one.  The subleading
operators are shown to be relevant to fine tune the S-wave decays
and improve the D-wave decays. Because of the rather large error
bars in and  significant correlations  between the resulting
effective coefficients, no clear conclusions about the physics
driving the $1/N_c$ corrections can be made.  The mixing angles
$\theta_1 = 0.39 \pm 0.11$ and $\theta_3 = (2.82, 2.38) \pm 0.11$
that result at NLO are similar to the ones determined at LO. They are ,
however,  somewhat different from the angles $\theta_1=0.61$ and
$\theta_3=3.04$ obtained in other analyses
\cite{CaroneGeorgi,CapstickRoberts,lll}.

\section*{ACKNOWLEDGEMENTS}

This work was supported by DOE contract DE-AC05-84ER40150 under which SURA operates the Thomas Jefferson
National Accelerator Facility,  by the National Science Foundation (USA) through grants \#~PHY-9733343
and -0300185 (JLG),  by CONICET (Argentina) under grant PIP 02368 and ANPCyT  (Argentina) under
grant PICT 00-03-08580 (CS and NNS), and
by the Fundaci\'on Antorchas (Argentina) (CS). The hospitality extended to the authors  by   the  $\rm ECT^*$ in Trento (Italy), where part of this work was completed,  is greatly appreciated.

\pagebreak

\begin{table}[htdp]
\caption{Negative parity non-strange baryons and their decay widths and branching ratios
from the PDG.
Channels not explicitly indicated are forbidden.}
\vspace*{1cm}\begin{center}
\begin{tabular}{ccccll} \hline \hline
\hspace*{.3cm}State \hspace*{.3cm}& \hspace*{.3cm}Notation \hspace*{.3cm}& \hspace*{.3cm}Mass \hspace*{.3cm}
& \hspace*{.3cm}Total width\hspace*{.3cm} & \multicolumn{2}{c}{Branching ratios [\%]} \\
    \cline{5-6}
  &&[MeV]&[MeV]& \hspace*{.3cm} S-wave\hspace*{.3cm} & \hspace*{.3cm} D-wave \hspace*{.3cm}\\ \hline
$N(1535)$& $N^*_{1/2}$  & $1538\pm 18$    & $150\pm 50 $ &
      $\pi N: 45\pm 10$              &     $\pi \Delta:0.5\pm 0.5$        \\
&&&&  $\eta N: 42.5\pm 12.5$         &                                    \\  \hline
$N(1520)$& $N^*_{3/2}$    & $1523\pm 8$   & $122 \pm 13$ &
  $\pi \Delta: 8.5\pm 3.5$           &        $\pi N:  55\pm  5$          \\
&&&&                                 &     $\pi \Delta: 12\pm 2$          \\
&&&&                                 &  $\eta N:~ {\rm unknown}$          \\  \hline
$N(1650)$ & $N^{*'}_{1/2}$ & $1660\pm 20$ & $167\pm 23$  &
      $\pi N: 72.5\pm 17.5$          &        $\pi \Delta: 4\pm 3$        \\
&&&&   $\eta N: 6.5\pm 3.5$          &                                    \\  \hline
$N(1700)$ & $N^{*'}_{3/2}$ & $1700\pm 50$ & $100\pm 50$ &
  $\pi \Delta: ~{\rm unknown}$       &   $\pi N:~      10\pm 5     $      \\
&&&&                                 &  $\pi \Delta: ~{\rm unknown}$      \\
&&&&                                 &     $\eta N:~ {\rm unknown}$       \\  \hline
$N(1675)$ & $N^*_{5/2}$ &  $1678 \pm 8 $ & $160\pm 20$  &
                                     &   $\pi N:~      45\pm 5$           \\
&&&&                                 &   $\pi \Delta: ~{\rm unknown}$     \\
&&&&                                 &       $\eta N:~ {\rm unknown}$     \\  \hline
$\Delta$(1620)&$\Delta^*_{1/2}$& $1645\pm 30$&  $150\pm 30$ &
  $\pi N:~ 25\pm 5$                  &   $\pi \Delta: 45\pm 15$           \\
&&&&           ~                     &       $\eta \Delta:
                                          \begin{array}{c}
                                   {\rm kinematically} \\ {\rm forbidden}
                                             \end{array} $                \\  \hline
$\Delta$(1700)&$\Delta^*_{3/2}$& $1720\pm 50$& $300\pm 100$ &
   $\pi \Delta: 37.5\pm 12.5$        &       $\pi N:~      15\pm 5$       \\
&&&&   $\eta \Delta:
        \begin{array}{c}
  {\rm kinematically} \\ {\rm forbidden}
       \end{array}$                  &   $\pi \Delta: 4\pm 3$             \\
&&&&                                 &     $\eta \Delta:
                                         \begin{array}{c}
                                   {\rm kinematically} \\ {\rm forbidden}
                                         \end{array}$                     \\ \hline \hline
\end{tabular}
\end{center}
\label{default}
\end{table}

\pagebreak

\begin{table}[h]
\begin{center}
\caption{Basis operators}
\vspace*{1cm}
\begin{tabular}{cccccccccc}\hline \hline
  & $n$-bodyness    &   Name  & Operator & Order in $1/N_c$ \\ \hline
    &  1B   &   $O_1^{[0,1]}$ &
$\left(\xi \ g\right)^{[0,1]}$ &
0 \\[2mm] \cline{2-5}
 Pion   &       &  $O_2^{[0,1]}$ &
$\frac{1}{N_c} \ \left(\xi \left(s\ T_c\right)^{[1,1]}\right)^{[0,1]}_{[0,a]}$
& 1 \\[2mm]
 S wave   &  2B   &  $O_3^{[0,1]}$ &
$\frac{1}{N_c} \ \left(\xi  \left(t\ S_c\right)^{[1,1]}\right)^{[0,1]}_{[0,a]}$
& 1 \\[2mm]
    &       &  $O_4^{[0,1]}$ &
$\frac{1}{N_c} \ \left(\xi \left(g \ S_c\right)^{[1,1]}\right)^{[0,1]}_{[0,a]}$
& 1 \\[2mm] \hline

     &   1B  & $O_1^{[2,1]}$ &
$\left(\xi \ g \right)^{[2,1]}_{[i,a]}$
& 0    \\[2mm] \cline{2-5}
     &       & $O_2^{[2,1]}$ &
$\frac{1}{N_c} \ \left(\xi \left(s \ T_c\right)^{[1,1]} \right)^{[2,1]}_{[i,a]}$
& $1$ \\[2mm]
 Pion    &       & $O_3^{[2,1]}$ &
$\frac{1}{N_c}  \ \left(\xi \left(t \ S_c\right)^{[1,1]} \right)^{[2,1]}_{[i,a]}$
& $1$ \\[2mm]
 D wave    &   2B  & $O_4^{[2,1]}$ &
$\frac{1}{N_c} \ \left(\xi \left(g \ S_c\right)^{[1,1]} \right)^{[2,1]}_{[i,a]}$
& $1$ \\[2mm]
     &     & $O_5^{[2,1]}$ &
$\frac{1}{N_c} \ \left(\xi  \left( g \ S_c \right)^{[2,1]} \right)^{[2,1]}_{[i,a]}$
& $1$ \\[2mm]
     &     & $O_6^{[2,1]}$ &
$\frac{1}{N_c} \ \left(\xi  \left( s \ G_c\right)^{[2,1]} \right)^{[2,1]}_{[i,a]}$
& $0$ \\[2mm] \cline{2-5}
     &  3B   & $O_7^{[2,1]}$ &
$ \frac{1}{N_c^2} \ \left(\xi
\left( s \left( \left\{ S_c, G_c \right\} \right)^{[2,1]}
\right)^{[2,1]}\right)^{[2,1]}_{[i,a]}$
& $1$ \\[2mm]
     &     & $O_8^{[2,1]}$ &
$\frac{1}{N_c^2} \ \left(\xi
\left( s \left( \left\{ S_c, G_c \right\} \right)^{[2,1]}
\right)^{[3,1]}\right)^{[2,1]}_{[i,a]}$
& $1$ \\[2mm] \hline
 Eta    &    1B  &  $O_1^{[0,0]}$ &
$\left( \xi \ s \right)^{[0,0]}_{[0,0]}$
& 0 \\[2mm] \cline{2-5}
S wave  &    2B  &  $O_2^{[0,0]}$ &
$\frac{1}{N_c} \left(\xi \left(s \ S_c\right)^{[1,0]} \right)^{[0,0]}_{[0,0]}$
& 1 \\[2mm] \hline
 Eta       &  1B   & $O_1^{[2,0]}$ &
$\left( \xi \  s \right)^{[2,0]}_{[i,0]}$
&  0 \\[2mm] \cline{2-5}
 D wave    &  2B   & $O_2^{[2,0]}$ &
$\frac{1}{N_c} \left(\xi \left(s \ S_c\right)^{[1,0]} \right)^{[2,0]}_{[i,0]}$
& 1  \\[2mm]
        &       & $O_3^{[2,0]}$ &
$\frac{1}{N_c} \left(\xi \left(s\ S_c\right)^{[2,0]} \right)^{[2,0]}_{[i,0]}$
& 1\\[2mm] \hline \hline
\end{tabular}
\end{center}
\end{table}

\pagebreak

\begin{table}[h]
\begin{center}
\caption{Reduced matrix elements of pion S wave operators}
\vspace{1cm}
\begin{tabular}{cccccc}
\hline \hline
Pion S waves  & \hspace{.5cm} $O^{[0,1]}_1$ \hspace{.5cm} & \hspace{.5cm} $O^{[0,1]}_2$ \hspace{.5cm}
& \hspace{.5cm} $O^{[0,1]}_3$ \hspace{.5cm} & \hspace{.5cm} $O^{[0,1]}_4$ \hspace{.5cm}  & \hspace{.5cm} Overall factor \hspace{.5cm} \\
\hline
   $^2N^*_{1/2}\rightarrow N$
         & $- 1$               & $-\frac{1}{2 N_c}$                   & $-\frac{1}{2 N_c}$       &
          $0$                  &
 $\frac{\sqrt{(N_c+3)(N_c-1)}}{3 N_c}$ \\
   $^2N^*_{3/2}\rightarrow  \Delta$
         & $\frac{1}{3}$       & $-\frac{1}{3 N_c}$                   & $-\frac{1}{3 N_c}$       &
          $\frac{1}{2\sqrt2 N_c}$    &
 $-\frac{\sqrt{3(N_c+5)(N_c+3)}}{\sqrt6 N_c}$\\
   $^4N^*_{1/2}\rightarrow  N$
         & $-\frac{1}{6}$      & $\frac{2}{3 N_c}$                    & $\frac{1}{6 N_c}$        &
           $\frac{1}{4\sqrt2 N_c}$   &
 $-\sqrt{2\frac{N_c-1}{N_c}}$ \\
   $^4N^*_{3/2}\rightarrow  \Delta$
         & $- \frac{1}{6}$     & $\frac{1}{6 N_c}$                    & $-\frac{1}{3 N_c}$       &
          $0$                  &
 $\sqrt{10 \frac{N_c+5}{N_c}}$\\
   $\Delta^*_{1/2}\rightarrow  N$
         & $\frac{1}{6}$       & $-\frac{1}{6 N_c}$                   & $-\frac{2}{3 N_c}$       &
           $0$                 &
 $-\sqrt{2 \frac{N_c-1}{N_c}}$\\
   $\Delta^*_{3/2}\rightarrow  \Delta$
         & $-\frac{1}{6}$      & $-\frac{1}{3 N_c}$                   & $\frac{1}{6 N_c}$        &
           $-\frac{1}{4\sqrt2 N_c}$  &
 $-\sqrt{10\frac{N_c+5}{N_c}}$\\  \hline
$\alpha^{[0,1]}$    & $\frac{3\sqrt3}{2\sqrt5}$   &
$\frac{3\sqrt3}{4\sqrt5}$ & $-\frac{3\sqrt3}{4\sqrt5}$  &
$\sqrt\frac{6}{5}$ &
\\ \hline \hline
\end{tabular}
\end{center}
\end{table}

\pagebreak
\begin{squeezetable}
\begin{turnpage}
\begin{table}[h]
\begin{center}
\caption{Reduced matrix elements of pion D wave operators}
\begin{tabular}{cccccccccc}\hline \hline
Pion D waves &\hspace{.5cm} $O^{[2,1]}_1$ \hspace{.5cm} &  \hspace{.5cm} $O^{[2,1]}_2$ \hspace{.5cm} & \hspace{.5cm} $O^{[2,1]}_3$ \hspace{.5cm}
& \hspace{.5cm} $O^{[2,1]}_4$ \hspace{.5cm} &
\hspace{.5cm} $O^{[2,1]}_5$ \hspace{.5cm} &  \hspace{.5cm} $O^{[2,1]}_6$ \hspace{.5cm} & $O^{[2,1]}_7$\hspace{.5cm}
&\hspace{.5cm} $O^{[2,1]}_8$ \hspace{.5cm} & \hspace{.5cm} Overall factor \hspace{.5cm} \\
\hline \hline
   $^2N^*_{1/2}\rightarrow \Delta$
& $-\frac{1}{3\sqrt 5}$  & $\frac{1}{3 \sqrt 5 N_c}$  & $\frac{1}{3\sqrt 5 N_c}$ & $- \frac{1}{2\sqrt{10} N_c}$  &
  $- \frac{1}{2\sqrt{30} N_c}$ &
 $\frac{N_c-1}{2\sqrt{30} N_c}$   & $\frac{1}{4\sqrt5 N_c^2}$ &
 0 & $5\frac{\sqrt{(N_c+3)(N_c+5)}}{2 N_c}$ \\[2mm]
   $^2N^*_{3/2}\rightarrow N$
& $-\frac{1}{\sqrt 5}$   & $-\frac{1}{2 \sqrt 5 N_c}$   & $-\frac{1}{2 \sqrt 5 N_c}$  &      0   &
 0  &
 0  & 0  &
 0 & $-5 \frac{\sqrt{(N_c+3)(N_c-1)}}{3  N_c}$
 \\[2mm]
   $^2N^*_{3/2}\rightarrow  \Delta$
& $-\frac{1}{3\sqrt 5}$  & $\frac{1}{3 \sqrt 5 N_c}$  & $\frac{1}{3\sqrt 5 N_c}$ & $-\frac{1}{2\sqrt{10} N_c}$  &
    $\frac{1}{2\sqrt{30} N_c}$  &
  $-\frac{N_c-1}{2\sqrt{30} N_c}$  &  $-\frac{1}{4\sqrt5 N_c^2}$ &
0 & $5 \frac{\sqrt{(N_c+3)(N_c+5)}}{2  N_c}$
\\[2mm]
   $^4N^*_{1/2}\rightarrow  \Delta$
& $-\frac{1}{6\sqrt 5}$  & $\frac{1}{6 \sqrt 5 N_c}$  & $-\frac{1}{3\sqrt 5 N_c}$  &      0             &
    $- \frac{1}{\sqrt{30} N_c}$  &
 $\frac{N_c-1}{4\sqrt{30} N_c}$  &  $-\frac{N_c-3}{4\sqrt5 N_c^2}$&
 $\frac{\sqrt{7}}{20}\frac{N_c - 1}{N_c^2}$ & $-5 \sqrt{\frac{N_c+5}{2 N_c}}$
\\[2mm]
   $^4N^*_{3/2}\rightarrow  N$
& $-\frac{1}{4\sqrt 5}$  & $\frac{1}{\sqrt 5 N_c}$    & $\frac{1}{4\sqrt 5 N_c}$ & $\frac{3}{8 \sqrt{10} N_c}$ &
     $- \frac{3\sqrt3}{8\sqrt{10} N_c}$     &
 $\frac{3\sqrt 3 (N_c + 2)}{8\sqrt{10}N_c}$ &  $-9\frac{N_c+1}{16\sqrt5 N_c^2}$ &
0 & $-\frac{2}{3}\sqrt{5\frac{N_c-1}{N_c}}$
\\[2mm]
   $^4N^*_{3/2}\rightarrow  \Delta$
& $-\frac{1}{3\sqrt 5}$  & $\frac{1}{3 \sqrt 5 N_c}$  & $-\frac{2}{3\sqrt 5 N_c}$  & 0 &
    $- \frac{1}{\sqrt{30} N_c}$   &
  $\frac{N_c-1}{4\sqrt{30} N_c}$  & $-\frac{N_c-3}{4\sqrt5 N_c^2}$  &
   $-\frac{\sqrt{7}}{40}\frac{N_c - 1}{N_c^2}$ & $-2\sqrt{5 \frac{N_c+5}{N_c}}$
\\[2mm]
   $N^*_{5/2}\rightarrow  N$
& $-\frac{1}{4\sqrt 5}$  & $\frac{1}{\sqrt 5 N_c}$    & $\frac{1}{4\sqrt 5 N_c}$ & $\frac{3}{8\sqrt{10} N_c}$ &
   $\frac{1}{8\sqrt{30} N_c}$        &
$-\frac{N_c+2}{8\sqrt{30} N_c}$   & $\frac{N_c+1}{16\sqrt5 N_c^2}$  &
0 & $- 2 \sqrt{5 \frac{N_c-1}{N_c}}$
\\[2mm]
   $N^*_{5/2}\rightarrow   \Delta$
& $\frac{1}{2\sqrt 5}$ & $-\frac{1}{2 \sqrt 5 N_c}$   & $\frac{1}{\sqrt 5 N_c}$   &      0             &
   $-\frac{1}{\sqrt{30} N_c}$ &
$\frac{N_c-1}{4\sqrt{30} N_c}$   &  $-\frac{N_c-3}{4\sqrt5 N_c^2}$
& $-\frac{N_c-1}{20\sqrt{7}N_c^2}$ & $5\sqrt{\frac{7}{10}
\frac{N_c+5}{N_c}}$
\\[2mm]
$\Delta^*_{1/2}\rightarrow  \Delta$
& $\frac{1}{6\sqrt 5}$ & $\frac{1}{3\sqrt 5 N_c}$  & $-\frac{1}{6\sqrt 5 N_c}$  & $\frac{1}{4\sqrt{10} N_c}$ &
    $\frac{1}{4\sqrt{30} N_c}$       &
$\frac{N_c+2}{20\sqrt{30} N_c}$      &  $\frac{3 N_c+1}{40\sqrt5
N_c^2}$ & 0 & $5 \sqrt{5 \frac{N_c+5}{N_c}}$
\\[2mm]
$\Delta^*_{3/2}\rightarrow   N$
& $-\frac{1}{6\sqrt 5}$  & $\frac{1}{6\sqrt 5 N_c}$  & $\frac{2}{3\sqrt 5 N_c}$ &      0             &
            0                      &
 0   &  0  &
0 & $-5 \sqrt{2 \frac{N_c-1}{N_c}}$
\\[2mm]
$\Delta^*_{3/2}\rightarrow  \Delta$
& $\frac{1}{6\sqrt 5}$ & $\frac{1}{3\sqrt 5 N_c}$  & $-\frac{1}{6\sqrt 5 N_c}$  & $\frac{1}{4\sqrt{10} N_c}$ &
 $-\frac{1}{4\sqrt{30} N_c}$     &
$-\frac{N_c+2}{20\sqrt{30} N_c}$ &   $-\frac{3 N_c+1}{40\sqrt5 N_c^2}$ &
0 & $ 5 \sqrt{5 \frac{N_c+5}{N_c}}$ \\ \hline
$\alpha^{[2,1]}$    &  $\sqrt{\frac{3}{7}}$   & $\frac{3\sqrt
3}{10\sqrt 2}$   & $\frac{\sqrt 3}{2\sqrt 7}$ & $\frac{2\sqrt
3}{5}$ & $\frac{3}{4}$ & $- \frac{12}{5}$ & $- \frac{3\sqrt6}{5}$
& $- \frac{3\sqrt3}{\sqrt{7}}$ &
\\ \hline \hline
\end{tabular}
\end{center}
\end{table}
\end{turnpage}
\end{squeezetable}
\pagebreak

\begin{table}[h]
\begin{center}
\caption{Reduced matrix elements of eta S wave operators}
\vspace*{1cm}
\begin{tabular}{cccc}
\hline \hline
Eta S waves  &\hspace{.5cm} $O^{[0,0]}_1$ \hspace{.5cm} & \hspace{.5cm} $O^{[0,0]}_2$ \hspace{.5cm}
& \hspace{.5cm} Overall factor \hspace{.5cm} \\
\hline
   $^2N^*_{1/2}\rightarrow  N$
         & $1$                 & $0$                      &   $-\frac{\sqrt{(N_c+3)(N_c-1)}}{\sqrt3 N_c}$ \\
   $^4N^*_{1/2}\rightarrow  N$
         & $1$                 & $-\frac{3}{2\sqrt2 N_c}$ &   $-\sqrt{2 \frac{N_c-1}{3 N_c}}$     \\
   $\Delta^*_{1/2}\rightarrow \Delta$
         & $1$                 & $\frac{3}{2\sqrt2 N_c}$ &    $\sqrt{2 \frac{N_c+5}{3 N_c}}$ \\
\hline
$\alpha^{[0,0]}$    & $\frac{3}{4}$   & $\frac{1}{\sqrt2}$   &
\\ \hline \hline
\end{tabular}
\end{center}
\end{table}

\begin{table}[h]
\begin{center}
\caption{Reduced matrix elements of eta D wave operators}
\vspace*{1cm}
\begin{tabular}{ccccc}\hline \hline
Eta D waves &\hspace{.5cm} $O^{[2,0]}_1$ \hspace{.5cm}   & \hspace{.5cm}  $O^{[2,0]}_2$ \hspace{.5cm}
& \hspace{.5cm}  $O^{[2,0]}_3$ \hspace{.5cm}  & \hspace{.5cm} Overall factor \hspace{.5cm} \\
\hline
   $^2N^*_{3/2}\rightarrow  N$
& $1$            & $0$         &        $0$
& $5 \frac{\sqrt{(N_c+3)(N_c-1)}}{\sqrt{15}  N_c}$ \\[2mm]
   $^4N^*_{3/2}\rightarrow  N$
& $1$            & $-\frac{3}{2\sqrt2 N_c}$  &  $\frac{3\sqrt3}{2\sqrt2 N_c}$
& $ -\sqrt{\frac{N_c-1}{3  N_c}}$ \\[2mm]
   $N^*_{5/2}\rightarrow  N$
& $1$            & $-\frac{3}{2\sqrt2 N_c}$ &  $-\frac{1}{2\sqrt6 N_c}$
& $-\sqrt{3 \frac{N_c-1}{N_c}}$ \\[2mm]
   $\Delta^*_{1/2}\rightarrow  \Delta$
& $1$            & $\frac{3}{2\sqrt2 N_c}$ & $\frac{\sqrt3}{2\sqrt2 N_c}$
& $5 \sqrt{\frac{N_c+5}{15 N_c}}$ \\[2mm]
   $\Delta^*_{3/2}\rightarrow   \Delta$
& $1$            & $\frac{3}{2\sqrt2 N_c}$ & $-\frac{\sqrt3}{2\sqrt2 N_c}$
& $5 \sqrt{\frac{N_c+5}{15 N_c}}$ \\[2mm] \hline
$\alpha^{[2,0]}$    &  $\frac{3}{2\sqrt{10}}$  & $\frac{1}{\sqrt{5}}$  &$-\sqrt{\frac{3}{5}}$  &
\\ \hline \hline
\end{tabular}
\end{center}
\end{table}

\pagebreak
\begin{table}[htdp]
\caption{ Fit parameters. Fit {\#1}: Pion S-waves LO.
In this case there is four-fold ambiguity for
the angles $\{ \theta_1 , \theta_3 \}$ given by the two values
shown for each angle. Fit {\#2}: Pion
S and D-waves, eta S-waves, LO. In this case there is two-fold ambiguity for
the angle $\theta_1$. For the angle $\theta_3$ there is an {\it almost} two-fold ambiguity
given by the two values indicated in parenthesis and which only differ in the two slightly
different values of $C^{[2,1]}_6 $. Fit {\#3}: Pion S and D-waves, eta S-waves,
NLO, no 3-body operators. No degeneracy in $\theta_1$  but
{\it almost} two-fold ambiguity in $\theta_3$ given by the two values indicated in parenthesis.
Values of coefficients which differ in the corresponding fits are indicated in parenthesis.} \vspace*{1cm}
\begin{center}
\begin{tabular}{cccc}
\hline \hline
Coefficient      &   \#1 LO     &   \#2 LO     &     \#3 NLO     \\  \hline
$C^{[0,1]}_1 $   &  $31 \pm 3 $ &   $31 \pm 3$           &   $  23\pm 3  $ \\
$C^{[0,1]}_2 $   &    -         &      -                 &   $ (7.4,32.5)   \pm (27,41) $ \\
$C^{[0,1]}_3 $   &    -         &      -                 &   $ (20.7,26.8)  \pm (12,14) $ \\
$C^{[0,1]}_4 $   &    -         &      -                 &   $ (-26.3,-66.8)\pm (39,65) $ \\  \hline
$C^{[2,1]}_1 $   &    -         & $4.6\pm 0.5$           &   $ 3.4\pm 0.3$ \\
$C^{[2,1]}_2 $   &    -         &      -                 &   $-4.5\pm 2.4$ \\
$C^{[2,1]}_3 $   &    -         &      -                 &   $ (-0.01,0.08)\pm 2$ \\
$C^{[2,1]}_4 $   &    -         &      -                 &   $ 5.7\pm 4.0$ \\
$C^{[2,1]}_5 $   &    -         &      -                 &   $ 3.0\pm 2.2$ \\
$C^{[2,1]}_6 $   &    -         & $(-1.86,-2.25)\pm 0.4$ &   $ -1.73\pm 0.26$ \\
$C^{[2,1]}_7 $   &    -         &      -                 &     -           \\
$C^{[2,1]}_8 $   &    -         &      -                 &     -           \\ \hline
$C^{[0,0]}_1 $   &    -         &  $11\pm 4$             &    $17\pm 4$           \\
$C^{[0,0]}_2 $   &    -         &      -                 &     -           \\ \hline
$\theta_1    $
                 & $ \begin{array} {c}1.62 \pm 0.12  \\ 0.29 \pm 0.11 \end{array} $
                 & $ \begin{array} {c}1.56 \pm 0.15  \\ 0.35 \pm 0.14 \end{array} $
                 & $ \begin{array} {c} 0.39 \pm 0.11 \end{array} $
                   \\ \hline
$\theta_3    $   & $ \begin{array} {c}3.01 \pm 0.07  \\ 2.44 \pm 0.06 \end{array} $
                 & $ \begin{array} {c}(3.00,2.44) \pm 0.07         \end{array} $
                 & $ \begin{array} {c}(2.82,2.38) \pm 0.11         \end{array} $
                   \\
\hline
$\chi^2_{\rm dof} $ &   0.25     &  1.5  &       0.9                   \\
${\rm dof}$         &     2     &  10   &       3                       \\ \hline \hline
\end{tabular}
\end{center}
\label{defaulta}
\end{table}

\pagebreak
\begin{table}[htdp]
\caption{ Partial widths resulting from the different fits in Table \ref{default}.
Values indicated in parenthesis correspond to the cases in which {\it almost} degenerate values of
$\theta_3$ lead to different partial widths.}
\vspace*{1cm}
\begin{center}
\begin{tabular}{cccccc}
\hline \hline
 \multicolumn{2}{c}{Decay} &\hspace{.5cm} Emp. Width   \hspace{.5cm}   &
 \hspace{.3cm} \#1 LO  \hspace{.3cm}   &   \hspace{.3cm} \#2 LO  \hspace{.3cm}
 &   \hspace{.3cm}  \#3 NLO   \hspace{.3cm}  \\
 & & [MeV] & [MeV] & [MeV] & [MeV] \\ \hline

  & $N(1535)\rightarrow \pi N$      & $68 \pm 27$  &
  74 &         62              &      (58,68)        \\

  & $N(1520)\rightarrow \pi \Delta$ & $10 \pm 4$   &
  10 &      9.7                 &       9.5       \\
\hspace{.5cm}$\pi$ S\hspace{.5cm}
  & $N(1650)\rightarrow \pi N$      & $121 \pm 40$ &
132  &     144                   &       122       \\
wave
  & $N(1700)\rightarrow \pi \Delta$ & unknown      &
 175 &     175                   &     (259,156)         \\

  & $\Delta(1620)\rightarrow \pi N$ &  $38 \pm 13$     &
  35 &      35                  &      38       \\

  & $\Delta(1700)\rightarrow \pi \Delta$ &  $112 \pm 53$     &
  81 &        81                &     (135,112)        \\
\hline

  & $N(1535)\rightarrow \pi \Delta$      & $1\pm 1$  &
     &           .01             &     0.5         \\

  & $N(1520)\rightarrow \pi    N $       & $67\pm 9$  &
     &          70              &       65       \\

  & $N(1520)\rightarrow \pi \Delta$      & $15\pm 3$  &
     &          2.8              &     13         \\

  & $N(1650)\rightarrow \pi \Delta$      & $7 \pm 5$  &
     &          0.12              &      8        \\
$\pi$ D
  & $N(1700)\rightarrow \pi N$           & $10 \pm 7$  &
     &         10               &      (11,9)        \\
wave
  & $N(1700)\rightarrow \pi \Delta$      & unknown  &
     &         4               &     (4,9)         \\

  & $N(1675)\rightarrow \pi N$           & $72 \pm 12$  &
     &         85               &      76        \\

  & $N(1675)\rightarrow \pi \Delta$      & unknown  &
     &        45                &      79        \\

  & $\Delta(1620)\rightarrow \pi \Delta$      & $68 \pm 26$  &
     &        30                &      87        \\

  & $\Delta(1700)\rightarrow \pi N$           & $45 \pm 21$  &
     &        49                &     32         \\


  & $\Delta(1700)\rightarrow \pi \Delta$      & $12 \pm 10$  &
     &        15                &      18        \\
\hline
$\eta$ S
  & $N(1535)\rightarrow \eta N$              & $64\pm 28$  &
     &        17                &     (57,61)         \\
wave
  & $N(1650)\rightarrow \eta N$              & $11\pm 6$  &
     &        14                &      12        \\ \hline
\end{tabular}
\end{center}
\label{resul}
\end{table}

\begin{thebibliography}{99}
\bibitem{foot}
There is, in addition, the pion
G-wave decay $N^*_{5/2} \rightarrow \pi \Delta_{3/2} $ which  is
not considered  here.
\end{thebibliography}
\end{document}